\documentclass [6pt,a4paper]{article}
\usepackage{bbm}
\usepackage{amsfonts}
\usepackage{mathrsfs}
\usepackage {amssymb}
\usepackage {amsmath}
\usepackage{amsthm}
\usepackage{latexsym}
\usepackage{extarrows}
\usepackage{chemarrow}
\def\dse#1{\vskip 0.6cm\noindent
        {\large\bf #1}
        \vskip 0.4cm}
\def\dse#1{\vskip 0.6cm\noindent
        {\large\bf #1}
        \vskip 0.4cm}
\usepackage{lastpage}
\usepackage{epsfig}

\begin{document}
\begin{center}
\textbf{\large{A note on the Singleton bounds for codes over finite
rings}}\footnote { E-mail addresses: ysh$_{-}$tang@163.com(Y. Tang),
693204301@qq.com(H. Xu), 1191398576@qq.com(Z.Sun).}
\end{center}

\begin{center}
{Yongsheng Tang$^{1}$,  Heqian Xu$^{1}$, Zhonghua Sun$^{2}$}
\end{center}

\begin{center}
\textit{\footnotesize $^{1}$School of Mathematics and Statistics,
Hefei Normal University, Hefei 230601, Anhui, P.R.China\\
$^{2}$School of Mathematics, Hefei University of Technology, Hefei
230009, Anhui, P.R.China}
\end{center}

\noindent\textbf{Abstract}~~In this paper,  we give a notation  on
the Singleton bounds  for linear  codes over a finite commutative
quasi-Frobenius ring in the work of Shiromoto [5]. We show that
there exists a class of finite commutative quasi-Frobenius rings.
The Singleton bounds  for linear  codes over such rings satisfy

\[
\frac{d(C)-1}{A}\leq n-\log_{|R|}|C|.
\]

\noindent\emph{keywords}: Linear codes, General weight,
Quasi-Frobenius ring, Singleton bounds

\dse{1~~Introduction}

Let $R$  be a finite commutative quasi-Frobenius (QF) ring  with
characteristic $k$ and  cardinality $M,$  where
$k=\prod_{i=1}^{s}p_{i}^{t_{i}},$
$M=\prod_{i=1}^{s}p_{i}^{r_{i}t_{i}},$ $p_{i}$ are distinct primes,
$r_{i}$ and $t_{i}$ are positive integers (see [4] and [7]). Let
$R^{n}$ be the free $R$-module of rank $n$ consisting of all
$n$-tuples of elements of $R.$  A linear code $C$ of length $n$ over
$R$ is a $R$-submodule of $R^{n}.$ An element of $C$ is called a
codeword of $C.$ In this paper, we will use a general notion of
weight, abstracted from the Hamming, the Lee and the Euclidean
weights. For each $c=(c_{1},c_{2},\cdots,c_{n})\in R^{n}$ and $r\in
R,$ the complete weight of $c$ is defined by
$$ n_{r}(c)=\{i |c_{i}= r\}.$$
To define a general weight function ${ w}(c)$, let $a_{r},$
$(0\neq)r\in R$ be positive real numbers, and set $a_{0}=0.$ Set
\begin{eqnarray}
{ w}(c)=\sum\limits_{r \in R}a_{r}n_{r}(c).
\end{eqnarray}
If we set $a_{r}=1,$ for every $(0\neq)r\in R,$ then ${w}(c)$ is
just the Hamming weight  of $c,$ denoted by ${w_{H} }(c).$
Throughout this paper, we denote
\begin{eqnarray}
A={\rm max}\{a_{r}|r\in R\}. \end{eqnarray} For a linear code $C$,
the general weight distance, denoted by $d(C)$, is equal to the
minimum general weight ${w}(C)$ of the code $C$.\\

\noindent\textbf{Example 1.1.}~~ Let us consider the ring
$R=\mathbb{Z}_{6}=\{0,1,2,3,4,5\}.$ If we set $a_{0}=0,
a_{1}=a_{5}=1, a_{2}=a_{4}=2$ and $ a_{3}=3,$ which yield the Lee
weights of the elements of $\mathbb{Z}_{6}$ (see [6]). Then $ A=3.$
While if we set $a_{0}=0, a_{1}=a_{5}=1, a_{2}=a_{4}=4$ and $
a_{3}=9,$ which yield the
Euclidean weights of the elements of $\mathbb{Z}_{6},$ then $A=9.$\\

\noindent\textbf{Example 1.2.}~~ Let us consider the ring
$\mathbb{F}_{2}+u\mathbb{F}_{2}=\{0,1,u,1+u\},$  which is a
2-dimensional algebra over the field $\mathbb{F}_{2}$ with a
nilpotent element $u$ where $u^{2}=0$ (see [1]). If we set $a_{0}=0,
a_{1}=a_{1+u}=1$ and $ a_{u}=2,$ which yield the Lee weights of the
elements of $\mathbb{F}_{2}+u\mathbb{F}_{2}.$ Then  $A=2.$ While if
we set $a_{0}=0, a_{1}=a_{1+u}=1$ and $ a_{u}=4,$  which yield the
Euclidean weights of the elements of $\mathbb{F}_{2}+u\mathbb{F}_{2},$ then  $A=4.$\\

 The following Singleton bounds on a general weight function
for linear code $C$
 over a finite commutative QF ring were obtained in [5]. \\

\noindent \textbf{Theorem 1.3.}\emph{ Let $C$ be a  linear code of
length $n$ over a finite commutative QF ring  $R.$ Let ${w}(c)$ be a
general weight function on $C$, as in (1); and with maximum
$a_{r}$-value $A$, as in (2). Suppose the minimum general weight of
${w}(c)$ on $C$ is $d(C)$. Then}

\[
\Big[\frac{d(C)-1}{A}\Big]\leq n-\log_{|R|}|C|,
\]
\emph{where $[b]$ is the integer part of $ b.$} \\

 The purpose of this
paper is to show that there exists a class of finite commutative QF
rings. The Singleton bounds  for linear  codes over such rings
satisfy

\[
\frac{d(C)-1}{A}\leq n-\log_{|R|}|C|.
\]

\dse{2~~General Gray map on $R$}
 Let $R$ be a finite commutative
quasi-Frobenius (QF) ring  with characteristic $k$ and  cardinality
$M,$  where $k=\prod_{i=1}^{s}p_{i}^{t_{i}},$
$M=\prod_{i=1}^{s}p_{i}^{r_{i}t_{i}},$ $p_{i}$ are distinct primes,
$r_{i}$ and $t_{i}$ are positive integers. For any element $a\in R,$
a general Gray map $\varphi$ on $R$ is defined as
$$\varphi : R  \rightarrow  \mathbb{F}_{p_{i}^{e_{i}}}^{A},$$
$$ a \mapsto  (a_{1},\ldots,a_{i},a_{i+1},\ldots,
a_{A}),$$ where $p_{i}^{e_{i}}$ is any divisor of
$\prod_{i=1}^{s}p_{i}^{r_{i}t_{i}}$   and
 $\mathbb{F}_{p_{i}^{e_{i}}}$ is a finite field with $p_{i}^{e_{i}}$ elements. In detail,
 \begin{itemize}
\item if $ a_{0}=0$,  ${w} (0)=0,$ then $\varphi(0)=(0,
\ldots,0, 0,\ldots, 0);$
\item if   $0 < a_{r}<A$ and ${w} (a_{r})=i,$ then $\varphi(a_{r})=(a_{1},\ldots,a_{i},a_{i+1},\ldots,
a_{A}),$  where there are  $i$  nonzeros and $A-i$ zeros among
$a_{1},\ldots,a_{i},a_{i+1},\ldots, a_{A};$
\item  if $ a_{r}=A$  and ${w}
(a_{r})=A,$ then $\varphi(a_{r})=(a_{1},\ldots,a_{i},a_{i+1},\ldots,
a_{A}),$ where $a_{t}\neq0$ for $ t=1,2, \ldots, A.$
 \end{itemize}
 The general Gray
map $\varphi$ can be extended to $R^{n}$ in an
obvious way.\\

\noindent\textbf{Example 2.1.}~~Let us consider a general  Gray map
$\varphi$ on $\mathbb{Z}_{6}$  with the Lee weight. From Example
1.1, we have $A=3.$ A general Gray map $\varphi$ from
$\mathbb{Z}_{6}$ to $\mathbb{F}_{m}^{3}$ ( $m=2$ or $3$), can be
defined as $\varphi(0)=(0,0,0),\varphi(1)=(0,0,
a_{11}),\varphi(2)=(0,a_{21},a_{22}),\varphi(3)=(a_{31},a_{32},a_{33}),
\varphi(4)=(a_{41},a_{42},0), \varphi(5)=(a_{51},0,0),$ where $a_{ij} \neq 0.$\\

From the definition
of the general Gray map we can obtain the following theorem.\\
\noindent{\bf Theorem 2.2.}~~\emph{Let the notation be as before.
For any  finite commutative QF ring $R,$ there exists a Gray map
$\varphi$ from $R^{n}$ to $\mathbb{F}_{p_{i}^{e_{i}}}^{An}$ and the
Gray map $\varphi$ is a distance preserving map from $(R^{n},$
general weight distance $)$ to $(\mathbb{F}_{p_{i}^{e_{i}}}^{An},$
Hamming distance $)$.}

 \noindent\textbf{Proof.} From the above
definitions, it is clear that $\varphi(x-y)=\varphi(x)-\varphi(y)$
for $x,y\in R^n$. Thus,
$d(x,y)=w(x-y)=w_{H}(\varphi(x-y))=w_{H}(\varphi(x)-\varphi(y))=d_{H}(\varphi(x),\varphi(y))$. \qed\\

\dse{3~~Main result}

In this section, we will  show that there exists a class of finite
commutative QF rings. The Singleton bounds  for linear  codes over
such rings satisfy

\[
\frac{d(C)-1}{A}\leq n-\log_{|R|}|C|.
\]
For our purpose, we firstly introduce the
following lemma on the Singleton bounds
for codes over finite fields (see [3]).\\

\noindent{\bf Lemma 3.1.}~~\emph{Let $C$ be a  code (possibly
nonlinear) of length $n$ over the finite field $\mathbb{F}_{q}$ with
$|C|$ elements and the minimum Hamming distance  $d_{H}(C)$.  Then}

\[
d_{H}(C) \leq n-\log_{q}|C|+1.
\]

\noindent\textbf{\upshape Theorem 3.2.} \emph{Let $C$ be a linear
code of length $n$ over a finite commutative QF ring  $R.$ Let
$w(x)$  be a general weight function on $C$, as in (1); and with
maximum $a_{r}$-value $A$, as in (2). Suppose the minimum weight of
$w(x)$ on $C$ is $d(C)$.  Let  $p=\emph{min}\{p_{1},\ldots, p_{s}
\}$ and $\varphi$ be a distance preserving map from $( R^{n},$
general weight distance) to $(\mathbb{F}_{p_{i}^{e_{i}}}^{An},$
Hamming distance).  Then }

\[
d(C)\leq An-\log_{p}|C| +1.
\]
\emph{Furthermore, if $\varphi$ is  a bijective map and a distance
preserving map from $( R^{n},$ general weight distance) to
$(\mathbb{F}_{p}^{An},$ Hamming distance), then}

\[
\frac{d(C)-1}{A}\leq n-\log_{|R|}|C|.
\]

\noindent \textbf{Proof.} By Theorem 2.2, we have each general Gray
image of the code $C$ under its corresponding general Gray map
$\varphi$ is a $p_{i}^{e_{i}}$-ary code with the same   length $An$
and the same minimum Hamming distance $d_{H}(\varphi(C)).$ By Lemma
3.1, each general Gray image of the code $C$ satisfies
\[ d_{H}(\varphi(C))\leq An-\log_{p_{i}^{e_{i}}}|\varphi(C)|+1.
\]
Furthermore

 \[
d(C)=d_{H}(\varphi(C))\leq \textrm{min}\{
An-\log_{p_{i}^{e_{i}}}|\varphi(C)|+1\}=An-\textrm{max}\{\log_{p_{i}^{e_{i}}}|\varphi(C)|
\}+1.
\]
 Since $p=\textrm{min}\{p_{1},\ldots, p_{s} \}$ and  $|\varphi(C)|\leq
 |C|,$ then
 $\textrm{max}\{\log_{p_{i}^{e_{i}}}|\varphi(C)|\}=\log_{p}|C|.$  Therefore

 \[
d(C)\leq An-\log_{p}|C| +1.
\]
 On the other hand,  if $\varphi$ is a bijection from $R^{n}$ to
$\mathbb{F}_{p}^{An}$, then $\log_{p_{i}^{t_{i}}}|\varphi(C)|$ is
maximum, that is,  if $\log_{p_{i}^{e_{i}}}|\varphi(C)|$ attains to
be maximum, then $|R|=p^{A}$ and $|\varphi(C)|=|C|. $  Hence, we
have
 \[
d(C)\leq An-\log_{p}|C| +1.
\]
It follows that

\[
\frac{d(C)-1}{A}\leq n-\log_{|R|}|C|.
\]
This completes the proof.
\qed\\

\noindent\textbf{Remark 3.2} Theorem 3.2 shows that if there exists
a bijective map $\varphi$   and the map $\varphi$ is  a distance
preserving map from $( R^{n},$ general weight distance) to
$(\mathbb{F}_{p}^{An},$ Hamming distance), then the Singleton bounds
for linear  codes over $R$  satisfy

\[
\frac{d(C)-1}{A}\leq n-\log_{|R|}|C|.
\]

\noindent \textbf{ \bf Example 3.3.}~~Consider any  linear code $C$
of length $n(\geq 1)$ over the QF ring
$\mathbb{F}_{2}+u\mathbb{F}_{2}=\{0,1,u,1+u\}.$  The Lee weights of
the elements of $\mathbb{F}_{2}+u\mathbb{F}_{2}$ is given as
follows:
 ${w}_{L}(0)=0, \ {w}_{L}(1)=1, \ {w}_{L}(u)=2, \ {w}_{L}(1+u)=1.$  For any element $ a + ub
\in \mathbb{F}_{2}+u\mathbb{F}_{2}. $ A Gray  map $\varphi$ from
$\mathbb{F}_{2}+u\mathbb{F}_{2}$ to $\mathbb{F}_{2}^{2}$ is defined
as  $\varphi(a+ub)=(b,a+b)$(see [1] ). The  map $\varphi$ can be
extended to $(\mathbb{F}_{2}+u\mathbb{F}_{2})^{n}$ in an obvious way
and the extended $\varphi$ is a bijection from
$(\mathbb{F}_{2}+u\mathbb{F}_{2})^{n}$ to $\mathbb{F}_{2}^{2n}.$
Therefore, we have

\[
\frac{d(C)-1}{2}\leq n-\log_{4}|C|.
\]

\noindent \textbf{ \bf Example 3.4.}~~Consider any  linear code $C$
of length $n(\geq 1)$ over the QF ring
$\mathbb{F}_{2}+u\mathbb{F}_{2}+v\mathbb{F}_{2}+uv\mathbb{F}_{2},$
which is a characteristic $2$  ring subject to the restrictions
$u^{2} = v^{2} = 0$  and $uv = vu.$ Let $\varphi:
(\mathbb{F}_{2}+u\mathbb{F}_{2}+v\mathbb{F}_{2}+uv\mathbb{F}_{2})^{n}
 \rightarrow \mathbb{F}_{2}^{4n}$ be the map given by
$\varphi(a + ub + vc + uvd) =(a + b + c + d, c + d, b + d, d).$ Then
$\varphi$ is a bijective map.  For any element $ a + ub + vc + uvd
\in \mathbb{F}_{2}+u\mathbb{F}_{2}+v\mathbb{F}_{2}+uv\mathbb{F}_{2},
$ we define the Lee weight $w_{L}(a + ub + vc + uvd) = w_{H}(a + b +
c + d, c + d, b + d, d), $ where $w_{H}$ denotes the ordinary
Hamming weight for binary codes(see [8]). Therefore, we have

\[
\frac{d(C)-1}{4}\leq n-\log_{16}|C|.
\]

\dse{4~~An application to codes over $\mathbb{Z}_{\ell}$}

In this section, we mainly  introduce the ring $\mathbb{Z}_{\ell}$
as a good example of a finite commutative QF ring.  Let
$\mathbb{Z}_{\ell}=\{0,1,,2, \ldots, \ell-1\} $ denote the ring of
integers modulo $\ell.$ Now we introduce three  kinds of weights,
namely the Hamming weight, the Lee weight, and the Euclidean weight.
The Hamming weight ${ w_{H} }(c)$ of a vector
$c=(c_{1},c_{2},\cdots,c_{n})\in \mathbb{Z}_{\ell}^{n},$ is the
number of its nonzero entries in the vector.  The Lee weight for the
elements of $\mathbb{Z}_{\ell}$ is defined as ${ w}_{{ L}} (a)={\rm
min}\{a, \ \ell-a\}$ for all $a\in\{0,1,\cdots,\ell-1\}$ (see [6]).
The Euclidean weight for the elements of $\mathbb{Z}_{\ell}$ is
defined as ${ w}_{ E} (a)={ w}_{ L} (a)^{2}$ for all
$a\in\{0,1,\cdots,\ell-1\}.$
 Denote the
minimum weight of a linear code $C$ with respect to Hamming, Lee,
Euclidean weights by $d_{H}(C)$; $d_{L}(C)$ and $d_{E}(C)$ ;
respectively. It is clear that the maximum $a_{r}$ -value is $1$;
$[\ell/2]$ and $[\ell/2]^{2}$; respectively. In the following,  we
denote by $\ell_{1}$ and $\ell_{2}$ the following integers,
respectively, $\ell_{1}=[\ell/2]$ and
$\ell_{2}=[\ell/2]^{2}.$\\

The next result follows immediately from Theorem 3.2.\\
\noindent\textbf{\upshape Corollary 4.1.} \emph{Let $C$ be a linear
code of length $n$ over over $\mathbb{Z}_{\ell},$  where
$\ell=\prod_{i=1}^{s}p_{i}^{a_{i}},$
 $p_{i}$ are distinct primes and $a_{i}$ are positive integers.
 Let  $p=\emph{min}\{p_{1},\ldots, p_{s}
\}.$   Then there are the following bounds on minimum weights: }

\[
d_{H}(C)\leq n-\log_{p}|C| +1;
\]
\[
d_{L}(C)\leq \ell_{1}n-\log_{p}|C|+1;
\]
\[
d_{E}(C)\leq \ell_{2}n-\log_{p}|C|+1.
\]
\emph{Furthermore, if $\varphi$ is  a bijective map and a distance
preserving map from $( \mathbb{Z}_{\ell}^{n},$ general weight
distance) to $(\mathbb{F}_{p}^{An},$ Hamming distance), then}
\[
d_{H}(C)-1\leq n-\log_{\ell}|C|;
\]
\[
\frac{d_{L}(C)-1}{\ell_{1}}\leq n-\log_{\ell}|C|;
\]

\[
\frac{d_{E}(C)-1}{\ell_{2}}\leq n-\log_{\ell} |C|.
\]

\noindent \textbf{ \bf Example 4.2.}~~Consider any  linear code $C$
of length $n(\geq 1)$ over the QF ring $\mathbb{Z}_{4}=\{0,1,2,3\}.$
The Lee weights of the elements of $\mathbb{Z}_{4}$ are given as
follows:
 ${w}_{L}(0)=0, \ {w}_{L}(1)={w}_{L}(3)=1, \ {w}_{L}(2)=2.$  Then $A=2.$  There exists a bijective map $\varphi$
from $\mathbb{Z}_{4}$ to $\mathbb{F}_{2}^{2}.$ In fact,
$\varphi(0)=(0,0)$, $\varphi(1)=(0,1)$, $\varphi(2)=(1,1)$, and
$\varphi(3)=(1,0)$(see [2] ). The  map $\varphi$ can be extended to
$\mathbb{Z}_{4}^{n}$ in an obvious way and the extended $\varphi$ is
a bijection from $\mathbb{Z}_{4}^{n}$ to $\mathbb{F}_{2}^{2n}.$
Therefore, we have

\[
\frac{d_{L}(C)-1}{2}\leq n-\log_{4}|C|.
\]
\dse{~~Acknowledgements} This research is supported by  Natural
Science Foundation of Anhui Province (No. 1408085QF116), Colleges
Outstanding Young Talents Program in 2014, Anhui Province ( No.
[2014]181) and Hefei Normal University Research Project (No.
2015JG09).


\begin{thebibliography}{99}

\bibitem{8} A. Bonnecaze, P. Udaya,  Cyclic codes and self-dual codes over $\mathbb{F}_{2}+u\mathbb{F}_{2}$, \emph{IEEE Trans. Inform. Theory} \textbf{4} (1999) 1250-1255.

\bibitem{2} A.R. Hammons, P.V. Kumar, A.R. Calderbank, N.J.A. Sloane, P. Sol\'{e}, The $\mathbb{Z}_{4}$-linearity of
Kerdock, Preparata, Goethals, and Related Codes, \emph{IEEE Trans.
Inform. Theory} \textbf{40} (1994) 301-319.


\bibitem{4} F.J. Macwilliams, N.J.A. Sloane,  \emph{The Theory of
Error-Correcting Codes}, North-Holland, Amsterdam New York, 1977.

\bibitem{2} B. R. McDonald, \emph{Finite Rings with Identity}, Dekker, New
York, 1974.


\bibitem{8}K. Shiromoto, Singleton bounds for codes over finite rings, \emph{Journal of Algebraic Combinatorics} \textbf{12} (2000) 95-99.


\bibitem{8} J.H. Van Lint,  \emph{Introduction to Coding Theory}, Third ed., Springer, Berlin, 1999.


\bibitem{12} J. Wood, Duality for modules over finite rings and applications to coding theory, \emph{American Journal of Mathematics}  \textbf{121} (1999) 555-575.


\bibitem{8}  B. Yildiz, S.Karadeniz, Linear codes over
$\mathbb{F}_{2}+u\mathbb{F}_{2}+v\mathbb{F}_{2}+uv\mathbb{F}_{2},$
\emph{Des. Codes Cryptogr.} \textbf{54} (2010)61-81.

\end{thebibliography}
\end{document}